\documentclass[final]{svjour2}
\usepackage{graphicx}
\usepackage{rotating}
\usepackage{amssymb}
\usepackage{mathptmx}
\usepackage[numbers]{natbib}
\makeatletter \journalname{Journal of Low Temperature Physics}

\bibpunct{}{}{,}{s}{}{,}

\begin{document}
\newcommand{\hdblarrow}{H\makebox[0.9ex][l]{$\downdownarrows$}-}
\title{Topological superfluid  $^3$He-B:  fermion zero modes on interfaces and in the vortex core}

\author{M.A. Silaev$^{1,2}$ and  G.E. Volovik$^{2,3*}$
}
\institute{
1:
Institute for Physics of Microstructures RAS, 603950 Nizhny Novgorog, Russia\\
\email{msilaev@mail.ru}
\\
2: Low Temperature Laboratory, Aalto University, P.O.
Box 15100, FI-00076 AALTO, Finland
\\ 3: L.D. Landau Institute for
Theoretical Physics, 119334
 Moscow, Russia
 \\ $^*$ Tel.: +358-9-4512963, Fax: +358-9-4512969, \email{volovik@boojum.hut.fi}
 }

\date{\today}

\maketitle


\begin{abstract}
Many quantum condensed matter systems are strongly correlated and
strongly interacting fermionic systems, which cannot be treated
perturbatively. However, topology allows us to determine generic
features of their fermionic spectrum, which are robust to perturbation and interaction.
We discuss the nodeless 3D system, such as superfluid $^3$He-B, vacuum of Dirac fermions,
and relativistic singlet and triplet supercondutors which may arise in quark matter.
The systems, which have nonzero value of topological invariant, have  gapless fermions
on the boundary and in the core of  quantized vortices.
We discuss the index theorem which relates fermion zero modes on vortices with the
topological invariants in combined momentum and coordinate space.


 \end{abstract}


\section{Introduction}

Fully gapped 3-dimensional fermionic systems -- topological insulator and topological supercondutors --
 are now under extensive investigation. The interest to such systems is revived after identification of
 topological insulators in compounds Bi$_{1-x}$Sb$_{x}$, Bi$_{2}$Te$_{3}$, Bi$_{2}$Se$_{3}$ (see review
 \cite{HasanKane2010}).  These systems are characterized by the gapless fermionic states on the boundary of the bulk insulator or at
   the interface between different states of the insulator. Historically, the topological insulators
    have been introduced in Ref. \cite{Volkov1981}, and  the first example of  fermion zero modes at the
    interface was provided in Ref. \cite{VolkovPankratov1985}.
     At the moment the term
`strong topological insulator' refers to time reversal invariant
insulators with odd number of Dirac points within the Fermi
surface on the surface of insulator, such as Bi$_{2}$Te$_{3}$
\cite{HasanKane2010,SCZhang2009}.
     The example of the fully gapped topological
    superfluids is superfluid $^3$He-B, discovered in 1972 \cite{OsheroffRichardsonLee1972}. The topological
     invariant for $^3$He-B and the gapless states at the interface between bulk states with different topological
      charges were discussed in Ref. \cite{SalomaaVolovik1988}. The modern theoretical treatment of topological
      insulators and superfluids/superconductors in  three spatial dimensions can be found in
       \cite{SCZhang2009b,Kitaev2009,Schnyder2008,Stone2010}.

Fully gapped 3-dimensional fermionic systems may arise also in
relativistic quantum field theories. In particular, the  Dirac
vacuum of massive Standard Model particles has also the nontrivial
topology, and the domain wall separating vacua with opposite signs
of the mass parameter $M$ contains fermion zero modes
\cite{JackiwRebbi1976}. Topologically nontrivial states may arise
in dense quark matter, where chiral and color superconductivity is
possible. The topological properties of such fermionic systems
have been recently discussed in Ref.  \cite{Nishida2010}. In
particular,  in some range of parameters the isotropic
 triplet relativistic superconductor is topological and has the fermion zero modes both at the boundary and in
  the vortex core. On the other hand, there is a range of parameters, where this  triplet superconductor is
  reduced to the non-relativistic superfluid $^3$He-B \cite{Ohsaku2001}. That is why
the analysis in Ref.  \cite{Nishida2010} is applicable to $^3$He-B
and becomes particularly useful when the fermions living in the
vortex core are discussed.

In relativistic theories there is an index theorem which relates the number of fermion zero modes localized
on a vortex with the vortex winding number \cite{JackiwRossi1981}. However, the analysis
in Ref.  \cite{Nishida2010} suggests, that this theorem is valid only for vortices in topological vacua
 (the Dirac vacuum considered in Ref.  \cite{JackiwRossi1981} is topological, see \cite{Volovik2010a}).
It is possible that there exists a more general index theorem
which relates  the number of fermion zero modes localized on a
vortex not only to the vortex winding number, but also to the
topological charge of the bulk vacuum or superconductor.

Here we consider the phase diagrams of the topologically different
states of isotropic triplet superconductors in relativistic regime
and in the non-relativistic weak coupling and strong coupling
regimes, and the fermion zero modes on domain walls and vortices
in these regimes. It appears that in all systems, which we
considered, a nonzero value of topological invariant in the bulk
system automatically leads to existence of  gapless fermions in
the core of  quantized vortices. We also demonstrate an example of
the index theorem, which expresses the number of fermion zero
modes on a vortex through the topological invariant in the
combined coordinate and momentum space.

\section{Superfluid relativistic medium and $^3$He-B}

 In relativistic superconductor or superfluid with the isotropic pairing -- such as  color superconductor in
  quark matter -- the fermionic spectrum is determined by Hamiltonian
  \begin{equation}
H=\tau_3\left( c {\mbox{\boldmath$\alpha$}}\cdot{\bf p} +  \beta M - \mu_R\right)+ \tau_1 \Delta
\,,
\label{eq:B-phaseRelatSinglet}
\end{equation}
for spin singlet pairing, and  by Hamiltonian
 \begin{equation}
H=\tau_3\left( c {\mbox{\boldmath$\alpha$}}\cdot{\bf p} +  \beta M - \mu_R\right)+ \gamma_5 \tau_1 \Delta
\,,
\label{eq:B-phaseRelat}
\end{equation}
for spin triplet pairing  \cite{Ohsaku2001,Nishida2010}.
Here $\alpha^i$, $\beta$ and $\gamma_5$ are Dirac matrices,
which in standard representation are
\begin{equation}
{\mbox{\boldmath$\alpha$}}  = \left( \begin{array}{cc}
0 & {\mbox{\boldmath$\sigma$}} \\
  {\mbox{\boldmath$\sigma$}} & 0 \end{array} \right)
  \;\;\;\; ,    \;\;\;\; \beta=\left( \begin{array}{cc}1 & 0 \\
   0 & -1 \end{array} \right)
    \;\;\;\; ,    \;\;\;\; \gamma_5=\left( \begin{array}{cc}0 & 1 \\
   1 & 0  \end{array} \right);
  \end{equation}
  $M$ is the rest energy of fermions; $\mu_R$ is their relativistic chemical potential as distinct
  from the non-relativistic chemical potential $\mu$;
  $\tau_a$ are matrices in Bogoliubov-Nambu space;
and $\Delta$ is the gap parameter.

 In non-relativistic limit the low-energy Hamiltonian is obtained by standard procedure, see e.g. \cite{Nishida2010b}.
  The non-relativistic limit is determined by the conditions
\begin{equation}\label{cond1}
 cp\ll M
\end{equation}
and
 \begin{equation}\label{cond2}
 |M-\sqrt{\mu_R^2+\Delta^2}|\ll M\,.
 \end{equation}
Under these conditions the Hamiltonian (\ref{eq:B-phaseRelatSinglet}) reduces to the
Bogoliubov - de Gennes (BdG)
Hamiltonian for fermions in spin-singlet $s$-wave superconductors, while (\ref{eq:B-phaseRelat})  transforms to the
BdG Hamiltonian relevant for fermions in isotropic spin-triplet $p$-wave superfluid $^3$He-B:
\begin{equation}
H=\tau_3\left(\frac{p^2}{2m} - \mu\right)+  c^B\tau_1{\mbox{\boldmath$\sigma$}} \cdot{\bf p}  ~~,~~ m=\frac{M}{c^2}~~,~~c^B=c\frac{\Delta}{M}
\,,
\label{eq:B-phase}
\end{equation}
 where the nonrelativistic chemical potential $\mu=\sqrt{\mu_R^2+\Delta^2}- M$. According to the
 Eq.(\ref{cond2}) the non-relativistic Hamiltonian is relevant
 only if $|\mu|\ll M$.
  The
Fermi liquid corrections are missing in this approach, which in
particular must give the effective mass
 $m^*$ instead of the bare mass $m$ of $^3$He atom. But this is not important for topological consideration.

 Note that we do not put any additional constraint on the value of the gap $\Delta$, which in principle
  can be comparable with the rest energy $M$, and as a result the velocity of particles
   $\partial \varepsilon/\partial p$ may approach the speed of light $c$ even in the non-relativistic
    limit $cp \ll M$. The special discussion is needed for the case when  $\Delta>M$.
     In this case one has $c_B>c$, and Eq.(\ref{eq:B-phase}) suggests that particles may
      propagate with velocity $|\partial \varepsilon/\partial p|>c$, which certainly is not correct
       since particles cannot move faster than light.  In this case of the large gap one should take
        into account the relativistic  corrections to the coefficient $c^B$ in Eq.(\ref{eq:B-phase}):
 \begin{equation}\label{Cb}
  c^B\approx \frac{\Delta}{M}\left(1-\frac{\mu}{2M}-\frac{c^2p^2}{4M^2}+\ldots\right).
 \end{equation}
 These corrections  together with  conditions (\ref{cond1}) and (\ref{cond2}) provide  the validity of
  equation $|\partial \varepsilon/\partial p|<c$ for the whole range of parameters.

 \begin{figure}
 \begin{center}
 \includegraphics[%
  width=0.8\linewidth,
  keepaspectratio]{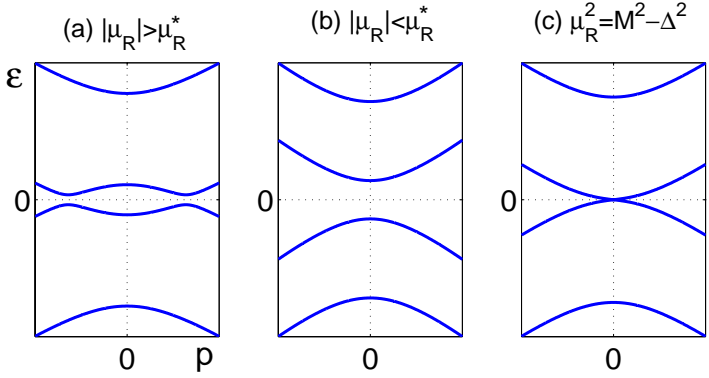}
\end{center}
 \caption{\label{PlotSpectrum} Plot of the spectrum of relativistic Hamiltonian
  (\ref{eq:B-phaseRelat})
 for two generic cases: (a) $|\mu_R| > \mu_R^*$ and (b)
 $|\mu_R| < \mu_R^*$. The gap in the spectrum closes at $\mu_R^2=M^2-\Delta^2$ which is shown in plot (c).}
\end{figure}

 The  Dirac-BdG system in Eq.(\ref{eq:B-phaseRelat}) has the
 following spectrum
 \begin{equation}\label{SpectrumGen}
  \varepsilon=\pm\sqrt{M^2+c^2p^2+\Delta^2+\mu_R^2\pm 2
  \sqrt{M^2(\mu_R^2+\Delta^2)+\mu_R^2c^2p^2}}.
 \end{equation}
 This spectrum is plotted in
Fig.\ref{PlotSpectrum}.
 Depending on the value of the parameters $\mu_R$, $\Delta$, $M$
 the spectral branches have different
 configurations.

There is a soft quantum phase transition, at which the
position of the minimum of energy $E(p)$ shifts from the origin ${\bf p}=0$,
and the energy profile forms the Mexican hat in momentum space.
This momentum-space analog of the Higgs transition  \cite{Volovik2007}
occurs when the relativistic chemical potential
  $\mu_R$ exceeds the critical value
  \begin{equation}
\mu^*_R=\left(\frac{M^2}{2} + \sqrt{ \frac{M^4}{4}
+M^2 \Delta^2}\right)^{1/2} \,. \label{eq:SoftTransition}
\end{equation}
 Fig.\ref{PlotSpectrum} demonstrates two generic cases:
 $|\mu_R| > \mu_R^*$
 when there are extremums of function $\varepsilon (p)$ at $p\neq 0$ and
  $|\mu_R| < \mu_R^*$ when all
  extremums are at the point $p=0$.

 \begin{figure}
 \begin{center}
 \includegraphics[%
  width=0.8\linewidth,
  keepaspectratio]{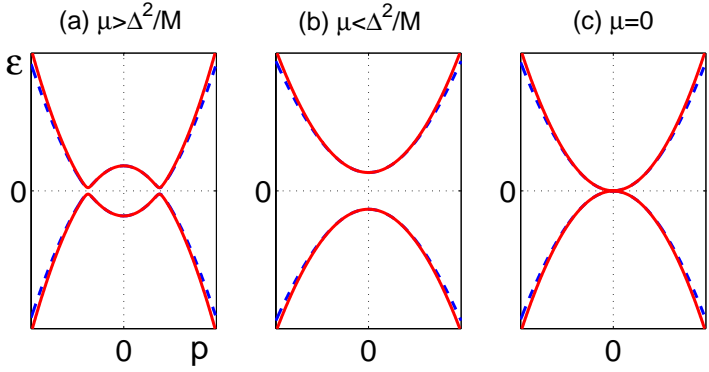}
\end{center}
  \caption{\label{PlotSpectrumNonRel}  Non-relativistic spectrum of
 Hamiltonian (\ref{eq:B-phase})
 (red solid lines) compared with the exact spectrum
 of relativistic Hamiltonian
  (\ref{eq:B-phaseRelat}) plotted by blue dash lines (only
 the lowest energy branches are shown).
 The two generic cases correspond to: (a) $\mu>\Delta^2/M$ and (b) $\mu<\Delta^2/M$.
 The gap in the spectrum closes at $\mu=0$ which is shown in plot (c).
 }
\end{figure}

  The spectrum of non-relativistic BdG Hamiltonian is plotted
  in Fig.\ref{PlotSpectrumNonRel}.
 In the non-relativistic limit the soft quantum transition
    takes place at $\mu^*=mc_B^2=\Delta^2/M$.
 But condition for derivation of the non-relativistic limit Eq.(\ref{cond2}) yields $| \mu | \ll M$. This condition
implies that the critical value $\mu^*$ is in the range of
applicability of non-relativistic limit only when $|\Delta| \ll
M$. For large $\Delta$ the Mexican hat is formed outside the
non-relativistic range.

The formation of the  Mexican hat at $|\mu_R| > \mu_R^*$ is an example of non-topological
 quantum phase transition.
Now we turn to the topological quantum phase transitions, at which the topological invariant changes.

\section{Topology of relativistic medium and  $^3$He-B}

 There is a characteristic line $\mu_R^2+\Delta^2=M^2 $ at which the gap in the spectrum closes  (see
  Fig.\ref{PlotSpectrum}c). In non-relativistic limit the node in the spectrum takes place at $\mu=0$ (see
  Fig.\ref{PlotSpectrumNonRel}c).
 This line corresponds to the topological quantum phase transition.  The vacuum states
  with $\mu_R^2+\Delta^2>M^2 $ and $\mu_R^2+\Delta^2<M^2$
 are characterized by different values of the topological invariant, and thus cannot be adiabatically connected.
  Discontinuity in the topological charge across the transition induces discontinuity in energy across the transition.
  For example, for the 2+1 $p_x+p_y$ superfluid/superconductor the phase transition is of third order,
  meaning that the third-order derivative of the ground state
energy is discontinuous \cite{Rombouts2010}.

 \begin{figure}
 \begin{center}
 \includegraphics[%
  width=0.6\linewidth,
  keepaspectratio]{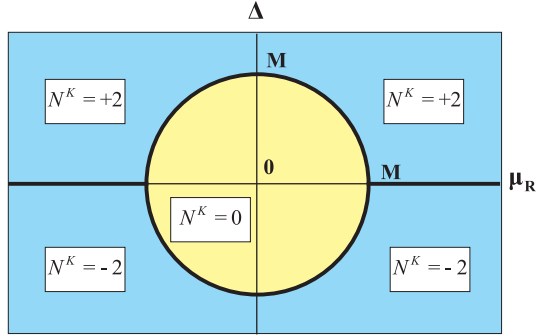}
\end{center}
  \caption{\label{PhaseDiagramRel}  Phase diagram of ground states of relativistic
 triplet superfluid in Eq.(\ref{eq:B-phaseRelat}) in the plane $(\mu_R, \Delta)$. Topological quantum phase transitions are marked by
  thick lines. The states inside the circle $\mu_R^2+\Delta^2=M^2$ are topologically trivial.
 The states outside this circle represent topological superconductors. The states on the lines of
  topological quantum phase transition are gapless.
 }
\end{figure}

The topological invariants which  describe the fully gapped superconductors/superfluids,
relativistic or non-relativistic, have the following form  \cite{Volovik2009a,Volovik2010a}
\begin{equation}
N^K = {e_{ijk}\over{24\pi^2}} ~
{\bf tr}\left[  \int   d^3p ~K
~H^{-1}\partial_{p_i} H
H^{-1}\partial_{p_j} H H^{-1}\partial_{p_k} H\right]\,,
\label{3DTopInvariant_tau}
\end{equation}
where the matrix $K$ reflects the symmetry of the system: it commutes or anti-commutes with the Hamiltonian.
The same invariants are applicable to the interacting systems,  but instead of Hamiltonian,
the Green's function matrix at zero
frequency must be used, $H({\bf p}) \rightarrow G^{-1}(\omega=0,{\bf p})$  \cite{Volovik2009a,Volovik2010a}.
 The  Green's function is the right object for the topological classification
of vacuum states, because it automatically takes into account interaction and works even in cases
 when the effective single-particle Hamiltonian is not available. In simple cases when the
 single-particle Hamiltonian can be introduced the invariants can be transformed to the forms
 proposed in the modern literature. For application of the Green's function for topological classification
 of gapless and fully gapped systems, see the book  \cite{Volovik2003} and review \cite{Volovik1004.0597}.

For $^3$He-B in Eq.(\ref{eq:B-phase}) and for triplet relativistic superconductor in Eq. (\ref{eq:B-phaseRelat})
the relevant matrix $K=\tau_2$. This matrix $K$, which anti-commutes with the Hamiltonian,
 is the combination of time reversal  and particle-hole symmetries.
Fig. \ref{PhaseDiagramRel}  shows the phase diagram of the vacuum states of relativistic  triplet superconductors.
 The states inside the circle $\mu_R^2+\Delta^2=M^2$ are topologically trivial, while
the states outside this circle represent topological superconductivity \cite{Nishida2010}. The states on the
 lines of topological quantum phase transition are gapless.

 \begin{figure}
 \begin{center}
 \includegraphics[%
  width=0.6\linewidth,
  keepaspectratio]{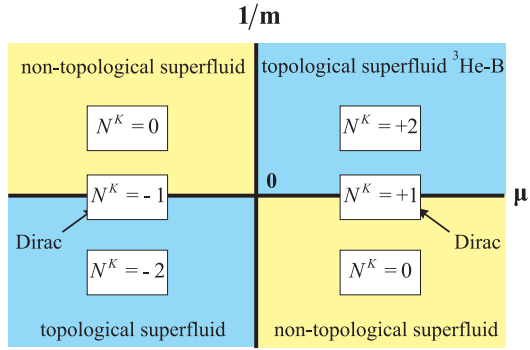}
\end{center}
  \caption{\label{3He-B}  Phase diagram of topological states of $^3$He-B in Eq.(\ref{eq:B-phase}) in the plane $(\mu,1/m)$. States on the line
  $1/m=0$ correspond to the  Dirac vacua, which Hamiltonian is non-compact. Topological charge of the Dirac fermions
  is intermediate between charges of compact $^3$He-B states.
The line $1/m=0$ separates the states with different asymptotic behavior of the Green's function at infinity:
$G^{-1}(\omega=0,{\bf p}) \rightarrow \pm \tau_3 p^2/2m$.
 The line $\mu=0$ marks topological quantum phase transition, which occurs between the weak coupling $^3$He-B
 (with $\mu>0$,
 $m>0$ and topological charge $N^K=2$) and the strong coupling $^3$He-B   (with $\mu<0$, $m>0$ and $N^K=0$).
  This transition is topologically equivalent to quantum phase transition between Dirac vacua with opposite mass parameter
 $M=\pm |\mu|$, which occurs when $\mu$ crosses zero along the line $1/m=0$.
 The interface which separates two states contains single Majorana fermion in case of $^3$He-B, and single chiral fermion
 in case of  relativistic quantum fields.  Difference in the nature of the fermions is that in Bogoliubov-de Gennes system
  the components of spinor are related by complex conjugation. This reduces the number of degrees of freedom compared
  to Dirac case.
 }
\end{figure}

Fig. \ref{3He-B} shows the phase diagram of the ground states of non-relativistic $^3$He-B in the plane $(\mu,1/m)$. The negative mass $m$ may appear in the band structure in crystals.
The topological quantum phase transition occurs at the critical value $\mu=0$,
which corresponds to the relativistic criterion $\mu_R= \sqrt{M^2-\Delta^2}$.
On the line  $1/m=0$ the Hamiltonian simulates that of  free Dirac fermions   with the mass parameter $M=\mu$ and the
 effective speed of light $c_{\rm eff}=c_B$. The vacuum of free Dirac vacuum  fermions has topological charge
\begin{equation}
N^K= {\rm sign}(M)
\,.
\label{eq:DiracInvariants}
\end{equation}

The real superfluid $^3$He-B lives in the weak-coupling corner of the phase diagram:
$\mu>0$, $m>0$, $mc^2\gg \mu\gg mc_B^2$.
However, in the ultracold Fermi gases with triplet pairing
 the strong coupling limit with $\mu <mc_B^2$ is possible near the Feshbach resonance \cite{GurarieRadzihovsky2007}.
 When $\mu$ crosses the value $\mu^{\rm c}=0$, the topological quantum phase transition occurs, at which the
  topological charge
 $N^K$ changes from  $N^K=2$ to  $N^K=0$.

 \begin{figure}
 \begin{center}
 \includegraphics[%
  width=0.6\linewidth,
  keepaspectratio]{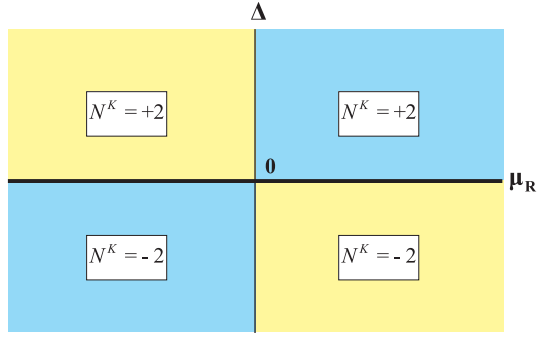}
\end{center}
  \caption{\label{PhaseDiagramRelSinglet}  Phase diagram of topological states of relativistic
singlet superfluid with $M=0$ in Eq.(\ref{eq:B-phaseRelatSinglet}) in the plane $(\mu_R, \Delta)$. }
\end{figure}

 The singlet relativistic superconductor may also have the nontrivial topology. This happens if  superconductivity
  occurs in the system of  massless fermions of Standard Model. The  equation (\ref{eq:B-phaseRelatSinglet})  with
  $M=0$, has additional symmetry. This symmetry leads to the matrix $K=\gamma_5\tau_2$, which anti-commutes with
  Hamiltonian (\ref{eq:B-phaseRelatSinglet}) at $M=0$. The phase diagram of states of the relativistic singlet
   superconductor at $M=0$ is shown in Fig. \ref{PhaseDiagramRelSinglet}.

\section{Gapless boundary states}

The simplified Hamiltonians describing the boundary states on the surface of topological superfluids or at the interface
between the bulk states:
 \begin{eqnarray}
\hat{H}=\tau_3\left( c {\mbox{\boldmath$\alpha$}}\cdot\hat{\bf p} +  \beta M - \mu_R\right)+ \gamma_5 \tau_1 \Delta(z)
\,,
\label{eq:B-phaseRelatBound}
\\
\hat{H}=\tau_3\left( c {\mbox{\boldmath$\alpha$}}\cdot\hat{\bf p}  - \mu_R\right)+  \tau_1 \Delta(z)
\,,
\label{eq:B-phaseRelatSingletBound}
\\
\hat{H}=\tau_3\left(\frac{\hat{\bf p}^2}{2m} - \mu\right)+   \tau_1 \left(c^B_x(z)\hat{p}_x\sigma_x+
c_y^B(z)\hat{p}_y\sigma_y+\frac{1}{2}\left\{c_z^B(z),\hat{p}_z\right\}\sigma_z\right),
\label{eq:B-phaseBound}
\\
\hat{H}= c\tau_3{\mbox{\boldmath$\sigma$}} \cdot\hat{\bf p}+
M(z)\tau_1\,. \label{eq:DiracBound}
\end{eqnarray}
Equations (\ref{eq:B-phaseRelatBound}) and  (\ref{eq:B-phaseRelatSingletBound}) describe the boundary/interfaces of
triplet and singlet relativistic superconductors correspondingly. The singlet relativistic superconductor is formed
by massless relativistic fermions, $M=0$. At the interfaces, the gap function $\Delta(z)$ changes sign.
Equation (\ref{eq:B-phaseBound}) is for the interfaces in superfluid $^3$He-B. At the interface, one or two or all
three components of speed $c^B$ change sign. The latter case corresponds
to the interface between two massive Dirac vacua in   Eq. (\ref{eq:DiracBound}), where the mass parameter $M(z)$
 changes sign across the interface.

\begin{figure}[top]
\centerline{\includegraphics[width=0.6\linewidth]{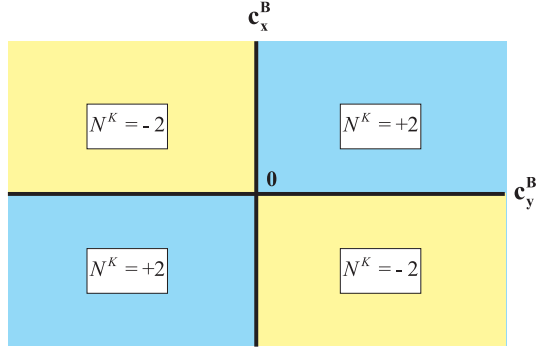}}
  \caption{\label{PDC}  Phase diagram of $^3$He-B states at fixed $c^B_z>0$, $\mu>0$ and $m>0$.
 At the phase boundaries the vacuum is gapless and corresponds to the 3+1 planar phase. The interface between the
  gapped states with different winding number $N^K$ contains Majorana fermions.
}
\end{figure}

 If the interface separates bulk states with different values of topological invariants $N^K$, such interface contains
  the fermion zero modes -- gapless branches of spectrum $E(p_x,p_y)$.
There is an index theorem which relates the number of fermion zero modes to the difference of topological invariants
of bulk states on two sides of the interface (see Refs. \cite{Volovik2003,Beri2010} and references therein).
 In $^3$He-B,
the topological charge changes sign if one of the speeds $c^B$  (see Fig.  \ref{PDC}) or all three speeds change sign.
 The latter case when all three speeds change sign across the interface is equivalent to the relativistic Ansatz in
  Eq.(\ref{eq:B-phaseRelatBound}). That is why the spectrum of fermion zero modes
at such wall can be obtained from applying the results of Ref. \cite{Nishida2010} to $^3$He-B. In the
 $^3$He-B limit  $mc_B^2\ll \mu$, one obtains the spectrum of fermion zero modes at small momentum:
  $E^2=v^2(p_x^2+p_y^2)$ with
velocity $v^2 \sim c_B^2\left(mc_B^2/\mu\right)$. This velocity is much smaller than the velocity $v\sim c_B$
of fermions localized at the interface at which only one of the three speeds, $c_z^B$ in Eq.(\ref{eq:B-phaseBound}),
 changes sign
\cite{Volovik2009b}. The latter interface mimics the boundary of $^3$He-B with specular reflection; the fermion
 zero modes at the boundary were discussed in Ref.  \cite{ChungZhang2009}.

\section{Fermion zero modes on vortices }

 The simplified Hamiltonians describing the fermionic  states in the core of a vortex with winding number
  $n$ are correspondingly:
 \begin{eqnarray}
\hat{H}=\tau_3\left( c {\mbox{\boldmath$\alpha$}}\cdot\hat{\bf p} +  \beta M - \mu_R\right)+ \gamma_5
\Delta(r)(\tau_1 \cos n\phi + \tau_2 \sin n\phi)
\,,
\label{eq:B-phaseRelatVortex}
\\
\hat{H}=\tau_3\left( c {\mbox{\boldmath$\alpha$}}\cdot\hat{\bf p}  - \mu_R\right)+   \Delta(r)(\tau_1
\cos n\phi + \tau_2 \sin n\phi)
\,,
\label{eq:B-phaseRelatSingletVortex}
\\
\hat{H}=\tau_3\left(\frac{\hat{\bf p}^2}{2m} - \mu\right)+   \frac{1}{2}\tau_1{\mbox{\boldmath$\sigma$}}
\cdot\left\{c^B(r)\cos n\phi,\hat{{\bf p}}\right\}+ \frac{1}{2}\tau_2{\mbox{\boldmath$\sigma$}}\cdot
\left\{c^B(r)\sin n\phi,\hat{{\bf p}}\right\},
\label{eq:B-phaseVortex}
\\
\hat{H}=  c\tau_3{\mbox{\boldmath$\sigma$}} \cdot\hat{\bf p} +M(r)  (\tau_1 \cos n\phi + \tau_2 \sin n\phi)
\,.
\label{eq:DiracVortex}
\end{eqnarray}
Eq.(\ref{eq:B-phaseVortex}) is the simplified Hamiltonian describing the most symmetric vortex in $^3$He-B:
even the simplest vortex -- most symmetric vortex with $n=1$, which is called the $o$-vortex, -- contains 5
components of the order parameter in the core \cite{SalomaaVolovik1987}.

Non-zero topological invariant describing the bulk superfluid gives rise to the gapless fermions living in the vortex core -- fermion zero modes. As distinct from the bound states at the interfaces, the general index theorem which relates the existence of the fermion zero modes to the topological charge of the bulk state and the vortex winding number is still missing. The existing index theorems are applicable only to particular cases. In relativistic systems the index theorem relates  the existence of the gapless fermions to the vortex winding number. There are also index theorems for fermions on vortices in non-relativistic systems: for the true fermion zero modes
\cite{Volovik1991} and for the Caroli-de Gennes-Matricon \cite{Caroli1964} spectrum which has a small gap (the so-called minigap) \cite{Volovik1993a}.
 For the general case, which takes into account both the momentum-space topology of bulk state and the real-space  topology of the vortex or other topological defects, the combined topology of the Green's function in the coordinate-momentum space $(\omega,{\bf p},{\bf r})$  \cite{GrinevichVolovik1988,Volovik1991,Volovik2003,TeoKane2009,TeoKane2010} must be used.
Here we consider examples, which demonstrate that the connection between the topological charge $N^K$ and the fermion zero modes on vortices. Other examples can be found in \cite{TeoKane2010,Herbut2010}.

 \begin{figure}
 \begin{center}
 \includegraphics[%
  width=0.8\linewidth,
  keepaspectratio]{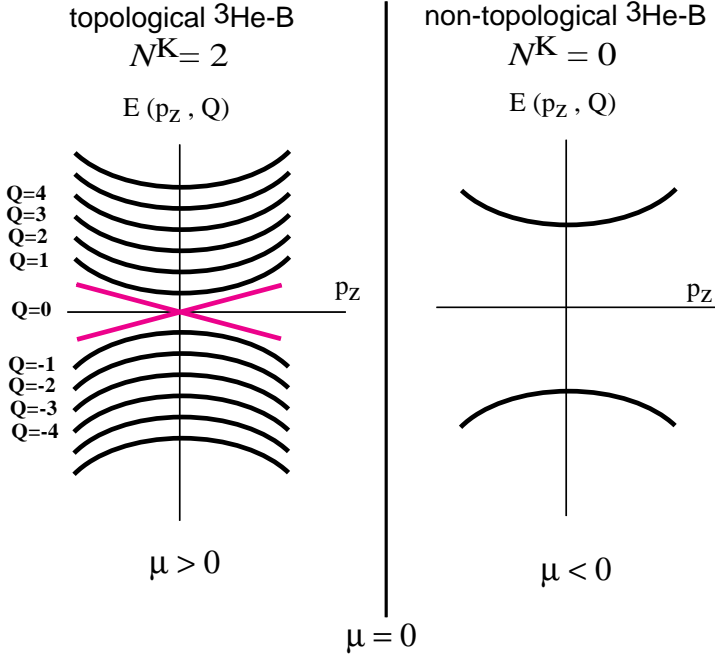}
\end{center}
  \caption{\label{FermionsBphaseVortex}
Schematic illustration of spectrum of  the fermionic bound states in the core
 of the most symmetric vortex with $n=1$, the so-called $o$-vortex \cite{SalomaaVolovik1987}, in fully gapped spin triplet superfluid/superconductor of $^3$He-B type.
 $Q$ is the azimuthal quantum number -- the generalized angular momentum of fermions in the vortex core.
({\it left}): Spectrum of bound state in the $^3$He-B, vortex which corresponds to the weak
coupling limit with non-zero topological charge $N^K=2$
\cite{MisirpashaevVolovik1995}.  There are two fermion zero modes, which cross zero energy in the opposite directions.  ({\it right}): The same vortex but in the
topologically trivial state of the liquid, $N^K=0$, does not have fermion zero modes. The spectrum
of bound states is fully gapped.  Fermion zero modes disappear at the topological quantum phase
transition, which occurs in bulk liquid at $\mu=0$. Similar situation may take place for strings
 in color superconductors in quark matter \cite{Nishida2010}.
 }
\end{figure}

 For $^3$He-B, which lives in the range of parameters where $N^K\neq 0$, the gapless fermions in the core have been
  found in Ref. \cite{MisirpashaevVolovik1995}. On the other hand, in the BEC limit, when $\mu$ is negative and the
  Bose condensate of molecules takes place, there are no gapless fermions. Thus in the BCS-BEC crossover  region the
  spectrum of fermions localized on vortices must be reconstructed.  The topological reconstruction of the fermionic
  spectrum in the vortex core cannot occur adiabatically. It should occur only during the topological quantum phase
  transition in bulk, when the bulk gapless state is crossed. Such topological transition occurs at $\mu=0$, see Fig.
   \ref{3He-B}.
At $\mu<0$  the topological charge $N^K$ nullifies and simultaneously the gap in the spectrum of core fermions arises,
 see Fig. \ref{FermionsBphaseVortex}. This is similar to the situation discussed in Ref. \cite{MizushimaMachida2010}
 for the other type of $p$-wave vortices, and in Refs. for Majorana fermions in semiconductor quantum wires
 \cite{Lutchyn2010,Oreg2010}.

This demonstrates that the existence of fermion zero modes is closely related to the topological properties of the
vacuum state. The reconstruction of the spectrum of fermion zero modes at the topological quantum phase transition
in bulk can be also seen for vortices in relativistic supercondutors  \cite{Nishida2010}. Let us first consider the
 triplet superconductor, which incorporates the $^3$He-B in the non-relativistic limit. The bulk states outside the
  circle  $\mu_R^2+\Delta^2=M^2$ in Fig. \ref{PhaseDiagramRel}  have   $N^K\neq 0$, and vortices in such bulk states
  do have fermion zero modes
\cite{Nishida2010}, while the states inside the circle are topologically trivial, and they have no zero modes in the
 vortex core \cite{Nishida2010}.

For singlet relativistic  superconductor, the topological invariant
$N^K$ with $K=\tau_2$ is always zero, and thus does not support the gapless fermions in the core.
However, at $M=0$, i.e. for chiral fermions there is another topological charge $N^K$ with $K=\tau_2\gamma_5$.
 This invariant is nonzero, see Fig. \ref{PhaseDiagramRelSinglet}, and this  is consistent with the existence of
  the fermion zero modes on vortices found by Nishida \cite{Nishida2010} for these superconductors at $M=0$.
  The fermion modes become gapped, when $M\neq 0$ and the topological invariant $N^K$ with $K=\tau_2\gamma_5$
   ceases to exist.

The generic example is provided by the  fermions on relativistic vortices  in Dirac vacuum
in Eq. (\ref{eq:DiracVortex}) discussed
in Ref. \cite{JackiwRossi1981}. The Dirac vacuum has the nonzero topological invariant, $N^K=\pm 1$, see Fig. \ref{3He-B}.
This is consistent with the existence of the fermion zero modes on vortices, found in Ref. \cite{JackiwRossi1981}.
The index theorem for fermion zero modes on these vortices can be derived using the topology in combined coordinate
and momentum space. Extending the results of  Ref. \cite{Volovik1991} for the spectral asymmetry index expressed via
the Green's function, one obtains that the algebraic number of zero modes -- branches which cross zero as function of
 $p_z$ -- is given by the 5-form  constructed from the Green's function:
\begin{eqnarray}
N_{\rm zm}=N_5(p_z \rightarrow +\infty)- N_5(p_z \rightarrow -\infty) \,,
\label{IndexTheorem}
\\
N_5(p_z) =
\nonumber
\\
\frac{1}{4\pi^3 i} ~
{\bf tr}\left[  \int   d^2p d^2x d\omega ~
G\partial_{p_x} G^{-1}
G\partial_{p_y} G^{-1}
G\partial_{x}G^{-1}
G\partial_{y} G^{-1}
G\partial_{\omega} G^{-1}
\right]\,.
\label{N5p_z}
\end{eqnarray}
In the simplest non-interacting case the Green's function is
$G^{-1}(\omega,{\bf p},x,y)=i\omega -H({\bf p},x,y)$.
 The 5-form invariant in terms of Green's function has been discussed also in \cite{Volovik2003,ZhongWang2010}.
 The Green's function
for Hamiltonians (\ref{eq:B-phaseRelatVortex})-(\ref{eq:B-phaseVortex}) has singularity at $x=y={\bf p}=\omega=0$,
and the integrals in (\ref{N5p_z}) are over two 5D planes, $p_z={\rm const}>0$ and $p_z={\rm const}<0$, on two sides
 of the singularity.

Choosing another
5D surface around the Green's function
singularity, the Eq.(\ref{IndexTheorem}) can be rewritten in the
following form:
 \begin{eqnarray}
 N_{\rm zm}=N_5 \,,
 \label{IndexTheorem2}
 \\
 N_5 = \nonumber
 \\
 \frac{1} {4\pi^3 i} ~
{\bf tr}\left[  \int   d^3p d\omega d\phi \,
 G\partial_{p_x}G^{-1}
 G\partial_{p_y} G^{-1}
 G\partial_{p_z}G^{-1}
 G\partial_{\omega} G^{-1}
 G\partial_{\phi} G^{-1}\right] \,.
 \label{N5}
 \end{eqnarray}
 The integral is now around the vortex line; the Green's function depends  on $(\omega,{\bf p}, r, \phi)$ with
  fixed distance $r>0$ from the vortex axis; and the azimuthal angle $\phi$ changes from $0$ to $2\pi$.

For the Dirac Hamiltonian (\ref{eq:DiracVortex}), equation (\ref{N5}) gives
$N_5=n$, which
 reproduces the index theorem discussed in Ref. \cite{JackiwRossi1981}: the algebraic number of fermion zero
 modes $ N_{\rm zm}$ equals the vortex winding number $n$. The integration over
the hyper-planes $p_z={\rm const}$
in Eq. (\ref{IndexTheorem}) yields the number which approaches $n$ in the asymptotic limit $|p_z|\gg |M|$.

 However, the Dirac vacuum is  marginal, since its Hamiltonian is non-compact \cite{Volovik2003,Schnyder2008}.
  The $^3$He-B provides the regularization of the Dirac vacuum at large momentum ${\bf p}$ \cite{Volovik2010a},
  so that the modified Dirac Hamiltonian becomes compact.
The modified Dirac vacuum with $1/m \neq 0$ in Fig. \ref{3He-B} is
either trivial ($N^K=0$) or topological ($N^K=\pm 2$).
 Correspondingly the fermion zero modes on a vortex either disappear  or are doubled,    Fig. \ref{FermionsBphaseVortex}.

The invariant $N_5$ does not give information on the fermion zero
modes for Hamiltonians
(\ref{eq:B-phaseRelatVortex})-(\ref{eq:B-phaseVortex}). For these
Hamiltonians one obtains $N_5=0$, which
 is consistent with Fig. \ref{FermionsBphaseVortex}: two branches have opposite signs of velocity $v_z$ and
 thus produce zero value for the algebraic sum of zero modes, $N_{\rm zm}=0$. To resolve the fermion zero modes in the systems
 where the branches cancel each other due to symmetry, the index theorem for the zero modes must be complemented
  by symmetry consideration.

\section{Conclusion}

We discussed here the fermion zero modes on vortices and interfaces which are protected by the topological
 invariant $N^K$.  In all the systems, which we considered,  the nonzero value of topological invariant in the
 bulk is associated with the existence of  gapless fermions
 in the core of  quantized vortices. The number of the fermion zero modes changes when the topological quantum
  phase transition occurs at which the bulk charge $N^K$ changes. This suggests that there must be the generalized
  index theorem which relates the number of fermion zero modes on the vortex with the topology in the combined
  coordinate and momentum $({\bf p},{\bf r})$ space. An example
 is provided by Eqs. (\ref{IndexTheorem}) and (\ref{IndexTheorem2}).

 \begin{figure}
 \begin{center}
 \includegraphics[%
  width=0.8\linewidth,
  keepaspectratio]{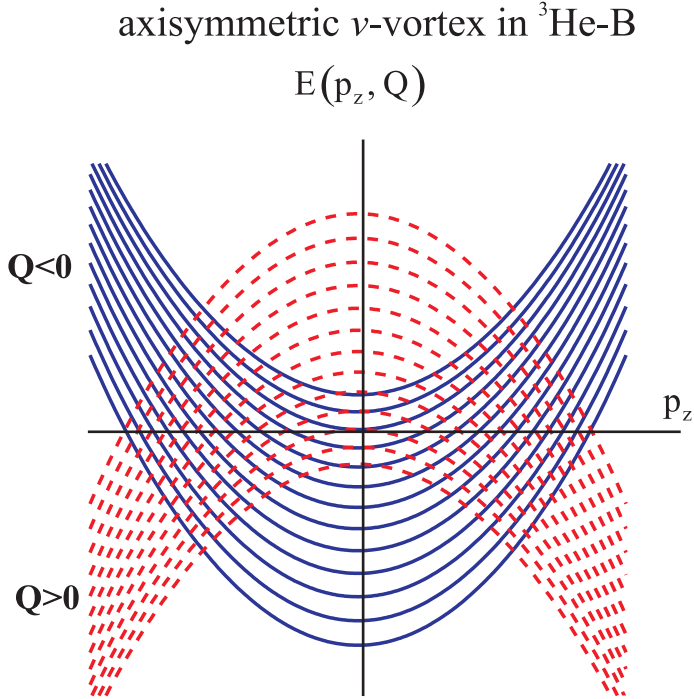}
\end{center}
  \caption{\label{v-vortex}
Schematic illustration of spectrum of  the fermionic bound states in the core
 of the real axisymmetric vortex in $^3$He-B (the $v$-vortex) \cite{Silaev2009}.
The fermion zero modes there belong to the class of one-dimensional Fermi surfaces.
Existence of these zero modes is supported by the 3-form topological invariant in terms of the Green's function in combined $({\bf p},{\bf r})$-space, $N_3 \propto {\bf tr}\int \left( G\partial G^{-1}\right)^3$, which gives rise to about $\left(\mu/mc_B^2\right)^{1/2}\sim 1000$
branches crossing zero energy  \cite{Volovik1991}.
 }
\end{figure}

We considered the most symmetric vortices. But in real superfluid $^3$He-B, the vortex cores have spontaneously
broken symmetry: the broken parity or the broken parity combined with broken axial symmetry \cite{SalomaaVolovik1987}.
  The broken parity leads to many branches of gapless fermions  forming the 1D Fermi liquids, as follows from the
  corresponding index theorem \cite{Volovik1991}.
Fig. \ref{v-vortex} illustrates these branches in one of the vortices observed in $^3$He-B -- in the axisymmetric
$v$-vortex with the spontaneously broken parity in the core \cite{Silaev2009}.
Each Fermi surface (the point where the branch $E_n(p_z)$ crosses zero), belongs to the class of nodes of
co-dimension 1 in Horava classification of topologically protected nodes \cite{Horava2005}.
These are the most stable objects in momentum space, they are protected by the topological invariant \cite{Volovik2003}
\begin{equation}
N_1={\bf tr}~\oint_C {dl\over 2\pi i}  G(p_0,{\bf p})\partial_l
G^{-1}(p_0,{\bf p})~.
\label{InvariantForFS}
\end{equation}
In general, the integral is taken over an arbitrary contour $C$ around the Green's function singularity
in  the $D+1$ momentum-frequency space $(i\omega,{\bf p})$; with $D=1$ for our case of one-dimensional
 Fermi liquids in the vortex core.
 Due to nontrivial topological invariant, the Fermi surface survives
 the perturbative interaction and exists even in case when quasiparticles are ill defined: in
 marginal, Luttinger and other exotic Fermi liquids  \cite{Faulkner2010}.

The topologically stable Fermi surface of co-dimension 1 may arise also on the surface of
 topological insulators forming the 2+1 Fermi liquids \cite{Fu2009}. In $^3$He-B, these
  topologically stable lines of nodes may arise at the interface at which two of three speeds
   $c^B$ change sign across the interface.  The bulk states on two sides of the interface have
   the same topological invariant, see Fig. \ref{PDC}, nevertheless the density of states of the
   fermion zero modes at such interface is non-zero at $E=0$
\cite{Volovik2009b}, which is the signature of 1D Fermi surface.

We discussed here the fermion zero modes on straight vortices, which appear in the
rotating vessel in superfluids or in applied magnetic field in superconductors.
However, for quantum turbulence phenomena \cite{Eltsov2009}, the dynamics of fermion zero
 modes within the entangled and curved vortices could become important, providing the possible
 source of dissipation of turbulence at very low temperature. Topology of the fermion zero modes on
  curved and entangled vortices has been  discussed in Ref. \cite{SCZhang2009b}.

\begin{acknowledgements}
It is a pleasure to thank Yuriy Makhlin for discussions. This work is supported in part by the Academy of Finland, Centers of
excellence program 2006-2011 and the Khalatnikov--Starobinsky leading scientific school (Grant No. 4899.2008.2). One of the authors (M.A.S.) was supported by ``Dynasty'' foundation.
\end{acknowledgements}

\pagebreak

\end{document}